\documentclass[sigconf]{acmart}
\usepackage{multirow}
\usepackage[english]{babel}
\usepackage[utf8x]{inputenc}
\usepackage{listings}
\AtBeginDocument{%
  \providecommand\BibTeX{{%
    \normalfont B\kern-0.5em{\scshape i\kern-0.25em b}\kern-0.8em\TeX}}}

\acmConference[Wiki Workshop 2022]{}{April 25, 2022}{online at The Web Conference 2022}
\begin{document}

\title[Reliability in Time: Evaluating the Web Sources of Information on COVID-19 in Wikipedia...]{Reliability in Time: Evaluating the Web Sources of Information on COVID-19 in Wikipedia across Various Language Editions from the Beginning of the Pandemic}

\author{Włodzimierz Lewoniewski}
\email{wlodzimierz.lewoniewski@ue.poznan.pl}
\orcid{0000-0002-0163-5492}
\affiliation{%
  \institution{\small{Poznan University of Economics and Business}}
  \streetaddress{Al. Niepodleglosci 10}
  \city{Poznan}
  \country{Poland}
  \postcode{61-875}
}

\author{Krzysztof Węcel}
\email{krzysztof.wecel@ue.poznan.pl}
\orcid{0000-0001-5641-3160}
\affiliation{%
	\institution{\small{Poznan University of Economics and Business}}
	\streetaddress{Al. Niepodleglosci 10}
	\city{Poznan}
	\country{Poland}
	\postcode{61-875}
}

\author{Witold Abramowicz}
\email{witold.abramowicz@ue.poznan.pl}
\orcid{0000-0001-5464-9698}
\affiliation{%
	\institution{\small{Poznan University of Economics and Business}}
	\streetaddress{Al. Niepodleglosci 10}
	\city{Poznan}
	\country{Poland}
	\postcode{61-875}
}

\renewcommand{\shortauthors}{W. Lewoniewski, K. Węcel and W. Abramowicz}

\begin{abstract}
There are over a billion websites on the Internet that can potentially serve as sources of information on various topics. One of the most popular examples of such an online source is Wikipedia. This public knowledge base is co-edited by millions of users from all over the world. Information in each language version of Wikipedia can be created and edited independently. Therefore, we can observe certain inconsistencies in the statements and facts described therein - depending on language and topic. In accordance with the Wikipedia content authoring guidelines, information in Wikipedia articles should be based on reliable, published sources. So, based on data from such a collaboratively edited encyclopedia, we should also be able to find important sources on specific topics. This effect can be potentially useful for people and organizations.

The reliability of a source in Wikipedia articles depends on the context. So the same source (website) may have various degrees of reliability in Wikipedia depending on topic and language version. Moreover, reliability of the same source can change over the time. The purpose of this study is to identify reliable sources on a specific topic – the COVID-19 pandemic. Such an analysis was carried out on real data from Wikipedia within selected language versions and within a selected time period.
\end{abstract}

\begin{CCSXML}
	<ccs2012>
	<concept>
	<concept_id>10002951.10003260.10003277.10003279</concept_id>
	<concept_desc>Information systems~Data extraction and integration</concept_desc>
	<concept_significance>500</concept_significance>
	</concept>
	<concept>
	<concept_id>10002951.10003227.10003233.10003301</concept_id>
	<concept_desc>Information systems~Wikis</concept_desc>
	<concept_significance>500</concept_significance>
	</concept>
	<concept>
	<concept_id>10003456.10003457.10003490.10003507.10003510</concept_id>
	<concept_desc>Social and professional topics~Quality assurance</concept_desc>
	<concept_significance>500</concept_significance>
	</concept>
	</ccs2012>
\end{CCSXML}

\ccsdesc[500]{Information systems~Data extraction and integration}
\ccsdesc[500]{Information systems~Wikis}
\ccsdesc[500]{Social and professional topics~Quality assurance}
\keywords{Wikipedia, references, COVID-19, information sources, reliability, information quality, Wikidata, DBpedia}

\maketitle

\section{Introduction}
High-quality information is essential for effective operation and decision-making in different types of organizations. This applies in particular to commercial companies, where the use of inaccurate and incomplete information may adversely affect their competitive advantage. Among over a billion different websites, very few have been a popular source of public knowledge for a relatively long time. Wikipedia is the largest encyclopedia ever created and is one of the popular open sources of multilingual information on the Web. Nowadays, this free encyclopedia has over 58 million articles written in over 320 languages \cite{url27}.

During the first few months of the COVID-19 pandemic, thousands of new Wikipedia articles on this topic were created and updated frequently by thousands of users. The high demand for information regarding the COVID-19 pandemic has resulted in a record number of views of these articles - hundreds of millions in a few months. Among such articles are those that provide the most important national statistics on COVID-19 cases, as well as the most important information about unfolding events related to the pandemic – whether regional, national or global. In order to provide high-quality data, the Wikipedia user community endeavors to ensure that reliable sources are sufficiently represented in the content of articles. This intent is to ensure that each represented fact in this collaborative encyclopedia can be checked by the reader. However, each language version may define its own criteria of source credibility, therefore, information about similar events in Wikipedia may have a different descriptions and references depending on the language. Moreover, these criteria may change over time, and, therefore, the reliability of some sources also changes. Since Wikipedia provides a history of edits to each article, it is possible to see each version of the page at a certain time and track which sources were reliable at any particular time.

The purpose of this study is to investigate important sources of information on the COVID-19 pandemic as provided in various Wikipedia languages. For this purpose, articles were identified that were thematically related to the subject of research. We showed how to solve this task by using the Wikipedia category system and semantic databases - Wikidata, DBpedia. In order to extract data about sources in Wikipedia articles in the different months, proprietary algorithms in Python were developed that took into account the complex structure of some articles.

\section{Reliable sources on Wikipedia}
Information on the Wikipedia should be based on reliable sources  \citep{url14}. Ideally, such sources should present all majority and significant minority views on some piece of information. This is important, as doing so ensures that readers of the article can be assured that each provided specific piece of information (statement) comes from a reliable and published source. Hence, before adding any information (even if it is a generally accepted truth) to this online encyclopedia, Wikipedia authors (volunteer editors or users) need to ascertain whether the facts put forward in the article can be verified by other readers \citep{url15}.

There is a wide range of works covering the field of sources analysis on Wikipedia. Some of the approaches use the number of the references to automatically assess quality of the information in Wikipedia  \citep{Stvilia2005,Blumenstock2008,Lucassen2010}. For example, external links (URLs) often appear in references where cited information is placed. Such links can be employed separately to assess quality of Wikipedia articles \citep{Yaari2011,Conti2014}. Additionally, such links in references can be assessed by indicating the degree to which these conform to their intended purpose \citep{tzekou2011}.

There are also studies analyzing the qualitative characteristics and metadata related to sources of Wikipedia articles. One of the works used special identifiers DOI and ISBN to unify the references and find the similarity of sources between various Wikipedia language editions \citep{lewoniewski2017ref}. Additionally we can find that a lot of sources in Wikipedia refers to scientific publications \citep{lewoniewski2017ref,nielsen2017scholia,lewoniewski2020,singh2021}. Such references often links to open-access works \citep{teplitskiy2017amplifying} and recently published journal articles \citep{jemielniak2019most}. Particularly popular are references about recent content, open access sources, life events \citep{piccardi2020}.

The availability of scientific sources makes Wikipedia especially valuable due to the potential of direct linking to other reliable sources. News websites are also one of the most popular sources of the information in Wikipedia and there is a method for automatic suggestion of the news references for the selected piece of information \citep{fetahu2016finding}. One of the research assessed a coverage of COVID-19-related scientific works cited in Wikipedia articles and found that information on this topic in Wikipedia comes from about 2\% of the scientific literature published at that time \cite{colavizza2020}. The aforementioned study also showed that editing users of Wikipedia are inclined to cite the latest scientific works and insert more recent information on the COVID-19 pandemic to Wikipedia shortly after the publication of these works.

But how do Wikipedia authors know which sources are reliable for use in Wikipedia articles? Basically, anyone intending to edit an article on Wikipedia needs to make that decision on his own. This is not trivial task, because reliability depends on topic and language version of Wikipedia. Each language can have own rules on how to find out if specific sources (websites, scientific journals etc.) can be considered appropriate for being providers of information. It is usually assumed that each editor can roughly tell if a source is reliable and can be used in a specific context. However, often such analysis is subjective, and there are no generally accepted quantitative measures to do so.

Only few developed language versions of Wikipedia contains non-exhaustive lists of sources whose reliability and use on Wikipedia are frequently discussed. For example the largest English Wikipedia \citep{url16} has such list with information on reliability for only about 300-400 sources. In that language edition of the encyclopedia we can also find such lists for specific topics (e.q. video games with about 600 sources \citep{url18}). Considering that currently the number of different websites is over billion \citep{url17,url19} and it is growing, a more complete list of such reliable sources of Wikipedia would be useful. Moreover, reliability of source in selected language version and topic can change with time - hence, such lists must be updated regularly.

\section{Automatic assessment of the Wikipedia sources} 

As it was mentioned before, presence of reliable sources affects the quality of Wikipedia articles. On the other hand, information with higher quality in Wikipedia must have appropriate references. So we can analyze sources placements to assess the reliability in context of topic and language. 

Different studies support that Wikipedia article length and number of references are important indicators for quality assessment of information \citep{Ferschke2012,Conti2014,Flekova2014,Dang2016a,Shen2017,Di2017}. Moreover, derived measures that are based on those indicators can improve existing quality models  \citep{Warncke-wang2013,Lewoniewski2017i,Lewoniewski2019comp}

Quality of Wikipedia articles depends also on quantity and experience of authors who contributed to the article. Often articles with the high quality are jointly created by a large number of different Wikipedia users, so such quantitative measure positively correlates with information quality in online encyclopedia \citep{Lih2004,Wilkinson2007,Kane2011,Liu2018,Lewoniewski2019comp}.

One of the recent works analyzed a behavior Wikipedia readers and found that overall engagement with citations is low and clicks from readers occur more frequently on Wikipedia articles of lower quality \citep{piccardi2020}. This work made a conclusion that about 1 in 300 page views results in a reference click. Other study showed that after loading a Wikipedia article, 0.8\% of the time reader hovers over a reference and 0.6\% clicks an external link \citep{wikiresearch1}.

Therefore, popularity can play an important role not only for quality estimation of information in specific language version of Wikipedia \citep{lewoniewskiphd} but also for checking reliability of the sources in it. Larger number of Wikipedia readers may allow for more rapid changes in incorrect or outdated information \citep{Lewoniewski2019comp}. Popularity of an article can be measured based on the number of page views or visits \citep{Lerner2018}. More over, popularity measures can be used not only for assessment of speciffic page, but also to measure the quality of entire websites \citep{bakaev2017}.

Reliability assessment of Wikipedia sources can be used in different approaches. One of the examples - integrating factual data of the best quality from various sources, such as Wikipedia, Wikidata, DBpedia and others \citep{hellmann2021}. Some of the important quality measures are implemented in some online services (such as \cite{url02} or \cite{url01}).

\section{Models for the web sources}
Recent study related to this work \citep{lewoniewski2020} proposed and implemented 10 models for reliability assessment of Wikipedia sources. Some of them also are implemented in special online web services (such as \cite{url03}). Such approaches uses measures that can be extracted from publicly available data (Wikimedia Downloads \citep{url20}), so anybody can use those models for different purposes. 

This work will use some of models that was proposed previously:

\begin{enumerate}
	\item \textbf{F} model -- based on frequency (F) of source usage. 
	\item \textbf{PR} model -- based on cumulative pageviews (P) of the article in which source appears divided by number of the references (R) in this article.
\end{enumerate}  

\textbf{F} model is one of the most basic and commonly used in relevant studies \citep{nielsen2007, lewoniewski2017ref, wikiresearch1, jemielniak2019most}. It assess how many times specific web domain occurs in external links of the references. For example, if the same source is cited 4 times, we count the frequency as 4. Equation \ref{eq1} shows the calculation for \textbf{F} model.

\begin{equation}
	F(s)=\sum_{i=1}^n C_s(i),
	\label{eq1} \end{equation}
where
$s$ is the source, 
$n$ is a number of the considered Wikipedia articles, 
$C_s(i)$ is a number of references using source $s$ (e.q. domain in URL) in article $i$.   

Quality of information in Wikipedia articles can be correlated with its page views \citep{Lerner2018,lewoniewskiphd,Lewoniewski2019comp}. This is primarily related to the fact that anyone can edit on Wikipedia, and that means if article was read by many people then more likely to have verified and reliable sources of information in it. In other words, the more readers can notice inappropriate source and there is bigger probability that one of such reader will make appropriate edit (to correct a source to more readable or to delete unverified information). 

\textbf{PR} model uses page views (visits) for certain period of time divided by the total number of the references in a considered Wikipedia article. Here visibility of the reference is also important. So, we can say that if more references are present in the article, then the less visible is a specific source for the particular reader (visitor). Equation \ref{eq2} shows the calculation using \textbf{PR} model.

\begin{equation}
	PR(s)=\sum_{i=1}^n \frac{V(i)}{C(i)} \cdot C_s(i),
	\label{eq2} \end{equation}
where
$s$ is the source, 
$n$ is a number of the considered Wikipedia articles, 
$C(i)$ is total number of the references in article $i$, 
$C_s(i)$ is a number of the references using source $s$ (e.q. domain in URL) in article $i$, 
$V(i)$ is page views (visits) value of article for certain period of time $i$.                                        

For purposes of this study, additionally \textbf{PR2} model will be used. It differs from \textbf{PR} model only by another way of counting page views - here only visits from humans will be taken into the account.

\section{Wikipedia articles related to COVID-19}
There different ways to obtain names of the Wikipedia articles on a specific topic. In following subsections three methods were presented: Wikipedia categories, Wikidata, DBpedia.

\subsection{Wikipedia categories}
There are diverse possibilities to find Wikipedia articles in different languages related to the COVID-19 pandemic. First of all, we can use information about categories aligned to the article. For instance, article ''COVID-19 pandemic in the United States'' in English Wikipedia \citep{url3} is assigned to such categories as: ''COVID-19 pandemic by country'', ''COVID-19 pandemic in the United States'', ''Presidency of Donald Trump'', ''Presidency of Joe Biden'', ''Trump administration controversies'', ''2020 in the United States'', ''2021 in the United States''. As we can see, not all of the categories are directly related to COVID-19 pandemic. Additionally, there is no single category that provide a full list of all Wikipedia articles related to COVID-19 pandemic in different places on the world. The most suitable and the most closest category to what we want to have is ''COVID-19 pandemic by country''. However, when we go to this category, we can only get directly list of Wikipedia articles which provide information about Coronavirus disease 2019 pandemic only for separate countries (as the name of the category suggests).

Additionally category ''COVID-19 pandemic by country'', we can also see links to other subcategories, which additionally have lists of Wikipedia articles that we need. For example, we can find there a category ''COVID-19 pandemic in India'' which contains a list of Wikipedia articles that potentially can be interesting for analysis. In particular, we will be able to find there such titles as: 
\begin{itemize}
	\item COVID-19 pandemic in India
	\item Timeline of the COVID-19 pandemic in India (January–May 2020)
	\item Timeline of the COVID-19 pandemic in India (June–December 2020)
	\item Timeline of the COVID-19 pandemic in India (2021)
\end{itemize}

However, category ''COVID-19 pandemic in India'' contains also Wikipedia articles, which are connected with pandemic, but not describes related events directly. For example, we can find separate articles that describes persons (e.q. who fight against the COVID-19 pandemic), video games, events (e.q. that were postponed) which are partially related to pandemic. Additional, in category ''COVID-19 pandemic in India'' we can find other subcategories that can be also considered to find articles on research topic. Here we can see links to following categories:
\begin{itemize}
	\item COVID-19 pandemic in India by state or union territory‎
	\item Deaths from the COVID-19 pandemic in India‎
	\item Impact of the COVID-19 pandemic in India‎
	\item Indian COVID-19 vaccines‎
	\item Timelines of the COVID-19 pandemic in India‎
\end{itemize}

It should be noted that there is no limitation on ''depth'' level in Wikipedia categories. Even more, links between categories are set by users, and they can only indicate the superior category for the current. Similar to Wikipedia articles, the categories are user-managed and anyone can make changes there. So, categories and links between them change dynamically in each language version of Wikipedia. As a consequence, we can encounter inconsistencies, tangles and other problems.

If, after all, we want to use the category system to identify all Wikipedia articles on the COVID-19 pandemic, we must find the main (the highest) category about this topic that allows us to generate the most complete list of such articles. Of course, as we have already seen with the example described above, to generate such list we also have to analyze links between categories.

It is not difficult to recognize that such a main category entitled ''COVID-19 pandemic''\citep{url2}. It contains subcategories that are relevant to research area: ''COVID-19 pandemic by location'', ''COVID-19 pandemic-related lists‎'', ''Statistics of the COVID-19 pandemic'' and other. However, not all subcategories will be taken into the analisys. For example category ''Deaths from the COVID-19 pandemic‎'' contains list of articles and other categories related to persons who whose cause of death was COVID-19. Wikipedia articles about those persons can include extensive information about their life and achievements, that are not connected with research topic of this paper (COVID-19 pandemic), therefore such information cannot be included in the analysis.

Structure of Wikipedia categories together with connection between articles and categories can be conducted through Wikipedia database backup dump files \citep{url4}. Three files have to be used (example for English Wikipedia):
\begin{itemize}
	\item \textbf{enwiki-latest-category.sql.gz} -- category information; here we use category identifiers and their names;
	\item \textbf{enwiki-latest-categorylinks.sql.gz} -- wiki category membership link records; here we use information about source page ID and destination category name;
	\item \textbf{en-latest-page.sql.gz} -- base per-page data; here we use pages ID, title and information about namespaces to identify articles (ns 0) and category (ns 14) pages.
\end{itemize}

There is also tool, that allow generate lists of Wikipedia articles based on categories titles - PetScan \citep{url5}.

\subsection{Wikidata} \label{sec:wikidata}
Wikidata is a semantic database that can be collaboratively edited by any interested person. Users can provide information to this knowledge graph using a web browser. Similarly to Wikipedia, Wikidata is a wiki service powered by the software: MediaWiki. Almost every article in Wikipedia has representation as Wikidata item. Moreover, Wikidata can be used to connect articles from different languages about the same subject - they will correspond to a single common Wikidata item that has its own unique identifier.

\begin{figure}[h]
	\centering
	\includegraphics[width=\linewidth]{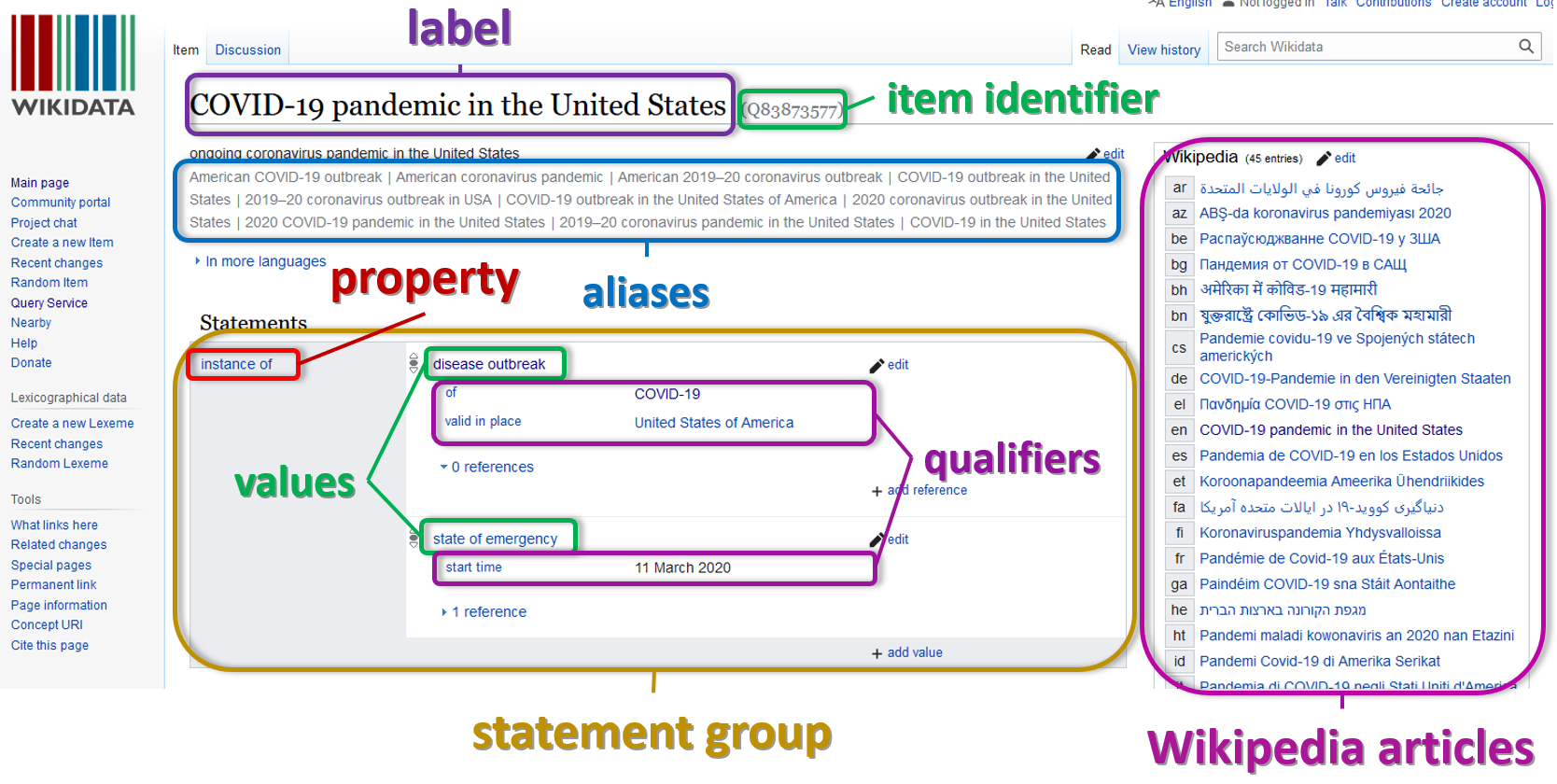}
	\caption{Scheme of the Wikidata item related to COVID-19 pandemic in the United States (Q83873577). Source: own work based on \cite{url1}.}
	\label{fig:wikidatascheme}
\end{figure}

Each Wikidata item has a collection of different statements structured in the form: ''Subject-Predicate-Object''. Figure \ref{fig:wikidatascheme}  shows Wikidata item Q83873577 with some statements.

Based on Wikidata statements we can find items on a specific topic. In our case, we will use the statement ''Property:P31 Q3241045'' (''instance of'' - ''decease outbreak'') with qualifier ''Property:P642 Q84263196'' (''of'' - ''COVID-19''). Listing \ref{lst:sparql} presents SPARQL query to get such list from Wikidata using its query service \citep{url6}. Result of this query is available on the web page: \url{https://w.wiki/48tx}.

\begin{lstlisting}[captionpos=b, caption=SPARQL query to get list of Wikidata items on disease outbreak of COVID-19, label=lst:sparql, basicstyle=\ttfamily,frame=single]
SELECT ?item WHERE {
	?item p:P31 [ps:P31 wd:Q3241045;
	pq:P642 wd:Q84263196]. }
\end{lstlisting}

Other important group of Wikidata items - timelines on COVID-19 pandemic. Listing \ref{lst:sparql2} presents SPARQL query to get such list from Wikidata using its query service \citep{url6}. Result of this query is available on the web page: \url{https://w.wiki/499G}.

\begin{lstlisting}[captionpos=b, caption=SPARQL query to get list of Wikidata items on Wikipedia timelines of COVID-19 pandemic, label=lst:sparql2, basicstyle=\ttfamily,frame=single]
SELECT ?item WHERE {
	?item p:P31 [ps:P31 wd:Q18340550;
	pq:P642 wd:Q81068910]. }
\end{lstlisting}

It is important to note that the presence of a Wikidata item does not mean the existence of a corresponding Wikipedia article in any one language version. Therefore, after obtaining the list of Wikidata items on COVID-19, we also need to obtain information about links to appropriate Wikipedia articles in selected languages.

\subsection{DBpedia}
One of the important part of Wikipedia articles is infoboxes, which present basic information about the subject in a convenient form. DBpedia is a semantic database, that extracts structured information from those infoboxes and other parts of Wikipedia articles, as well as information extraction from other Wikimedia projects \citep{lehmann2015}. Such knowledge is extracted from different languages versions of Wikipedia articles to the form of semantic triples (''Subject-Predicate-Object'') in unified form using DBpedia ontology.

There are specific infoboxes that may indicate COVID-19 pandemic related subjects (Wikipedia articles). For example, ''Infobox pandemic'' (which redirects to ''Infobox outbreak'') used on the high-profile COVID-19 articles \citep{url7}. However, such infobox can be also transcended in articles related to other pandemics (such as Spanish flu, 2009 swine flu pandemic, 2002–2004 SARS outbreak and others). To select only COVID-19 related Wikipedia articles we also need to consider specific parameter of the infobox - ''disease''. DBpedia can help to identify such Wikipedia infoboxes in various wordings and their parameters depending on language version. 

\begin{figure}[h]
	\centering
	\includegraphics[width=\linewidth]{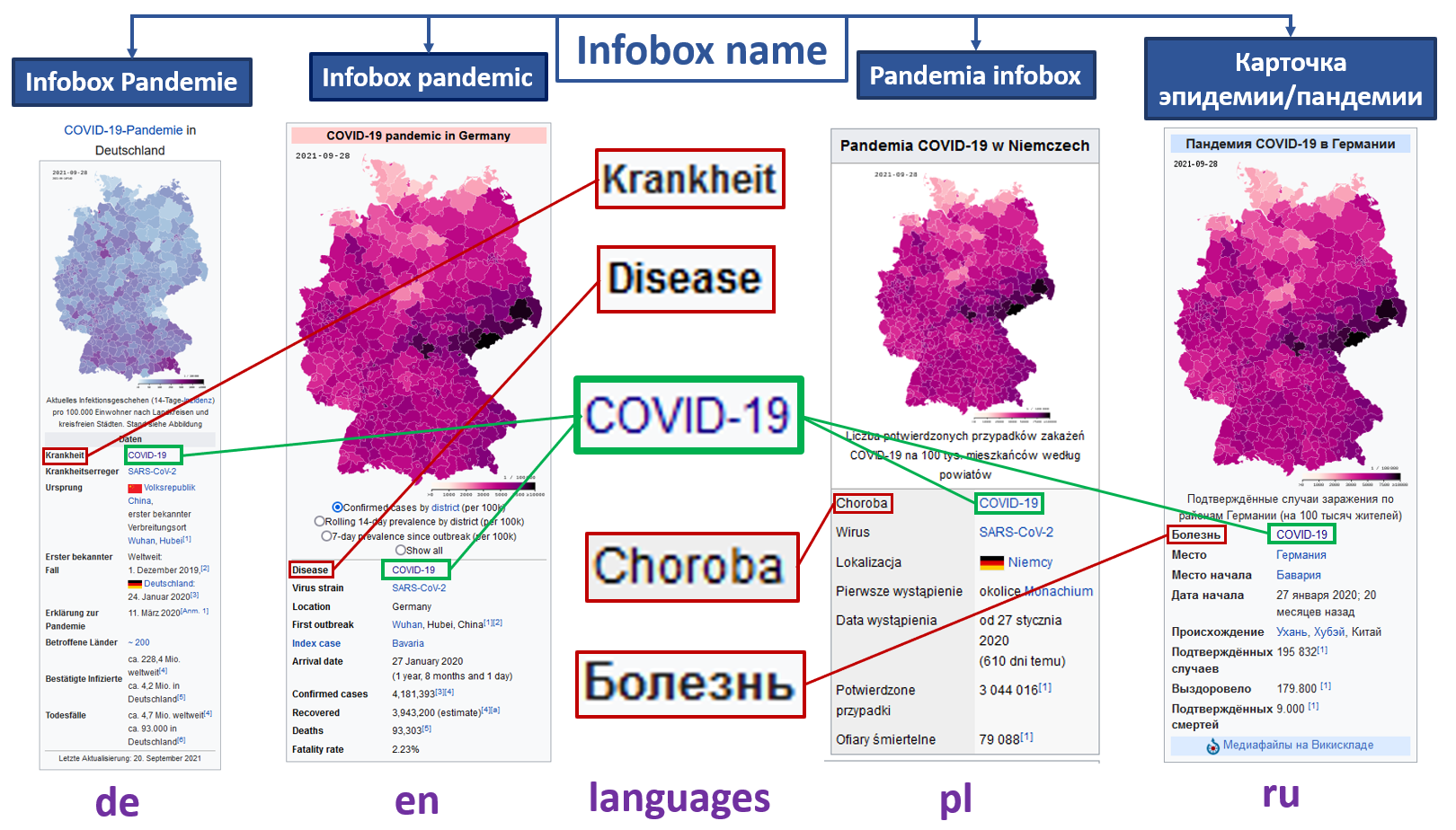}
	\caption{Infoboxes on COVID-19 pandemic in Germany in four language editions of Wikipedia. Source: own work based on \cite{url9,url8,url10,url11}}
	\label{fig:infoboxes}
\end{figure}

Figure \ref{fig:infoboxes} illustrates infoboxes on COVID-19 pandemic in Germany from 4 language editions of Wikipedia.

After extraction, DBpedia created separate pages for each Wikipedia article in selected language. Such page on DBpedia contains structured knowledge related to subject. For example, there is resource ''COVID-19 pandemic in Germany'' on DBpedia \citep{url12}, which is a result of information extracting from corresponding article in English Wikipedia about this topic \citep{url8}.

Similarly to Wikidata (described in previous subsection), DBpedia also have own online query editor \citep{url13} which can be used to generate list of resources with specific statements.

\subsection{Selected Wikipedia articles on COVID-19}
For purpose of this study 15 the most developed language versions (or chapters) of Wikipedia were selected. Those chapters contained at least 1,000,000 articles and had over 10 value of depth indicator (showing how frequently its articles are updated) as of September 2021. Table \ref{tab:articlescount} presents those Wikipedia languages with number of all articles and identified as related to COVID-19 pandemic.

\begin{table}[htbp]
	\centering
	\begin{scriptsize}
		\begin{tabular}{|l|r|r|}
			\cline{1-3}
			\multicolumn{1}{|c|}{\multirow{2}{*}{\textbf{Languages}}} & \multicolumn{2}{c|}{\textbf{Number of articles}}           \\ \cline{2-3}
			\multicolumn{1}{|c|}{}                                    & \multicolumn{1}{c|}{\textbf{All}} & \multicolumn{1}{c|}{\textbf{COVID-19 pandemic}}  \\ \cline{1-3}
			ar - Arabic                                               & 1,133,676                         & 371                                              \\ \cline{1-3}
			de - German                                               & 2,610,474                         & 237                                              \\ \cline{1-3}
			en - English                                              & 6,368,182                         & 547                                              \\ \cline{1-3}
			es - Spanish                                              & 1,711,182                         & 276                                              \\ \cline{1-3}
			fr - French                                               & 2,357,100                         & 208                                              \\ \cline{1-3}
			it - Italian                                              & 1,714,274                         & 184                                               \\ \cline{1-3}
			ja - Japanese                                             & 1,287,349                         & 51                                               \\ \cline{1-3}
			nl - Dutch                                                & 2,065,186                         & 46                                               \\ \cline{1-3}
			pl - Polish                                               & 1,487,980                         & 41                                                \\ \cline{1-3}
			pt - Portuguese                                           & 1,074,240                         & 255                                               \\ \cline{1-3}
			ru - Russian                                              & 1,750,351                         & 129                                               \\ \cline{1-3}
			sv - Swedish                                              & 2,945,171                         & 7                                                \\ \cline{1-3}
			uk - Ukrainian                                            & 1,111,954                         & 254                                               \\ \cline{1-3}
			vi - Vietnamese                                           & 1,268,830                         & 211                                               \\ \cline{1-3}
			zh - Chinese                                              & 1,225,098                         & 218                                               \\ \cline{1-3}
		\end{tabular}
	\end{scriptsize}
	\caption{Selected language versions of Wikipedia with information about number of all articles and related to COVID-19.}
	\label{tab:articlescount}
\end{table}

%PLUS
%COVID-19 pandemic by location‎ 
%COVID-19 pandemic-related lists‎ 
%%Timelines of the COVID-19 pandemic‎
%Responses to the COVID-19 pandemic
%Statistics of the COVID-19 pandemic

%MINUS
%COVID-19 conspiracy theorists‎
%Deaths from the COVID-19 pandemic
%Impact of the COVID-19 pandemic‎
%COVID-19 pandemic in popular culture‎

%Category:WikiProject COVID-19 articles
%https://petscan.wmflabs.org/

\section{Data extraction from revision history of Wikipedia articles on COVID-19 pandemic}

\subsection{References extraction}
After identifying the Wikipedia articles related to research topic, we need to analyze their edition history to know what content and sources they had in particular day. One of the possibility is to find selected articles in dumps with complete page edit history. Another way - is to get such data from dedicated Wikiped API service. However, in such approach we need to request data for each article separately, and for articles with a large number of editions - several requests for the same Wikipedia article.

\begin{figure}[h]
	\centering
	\includegraphics[width=\linewidth]{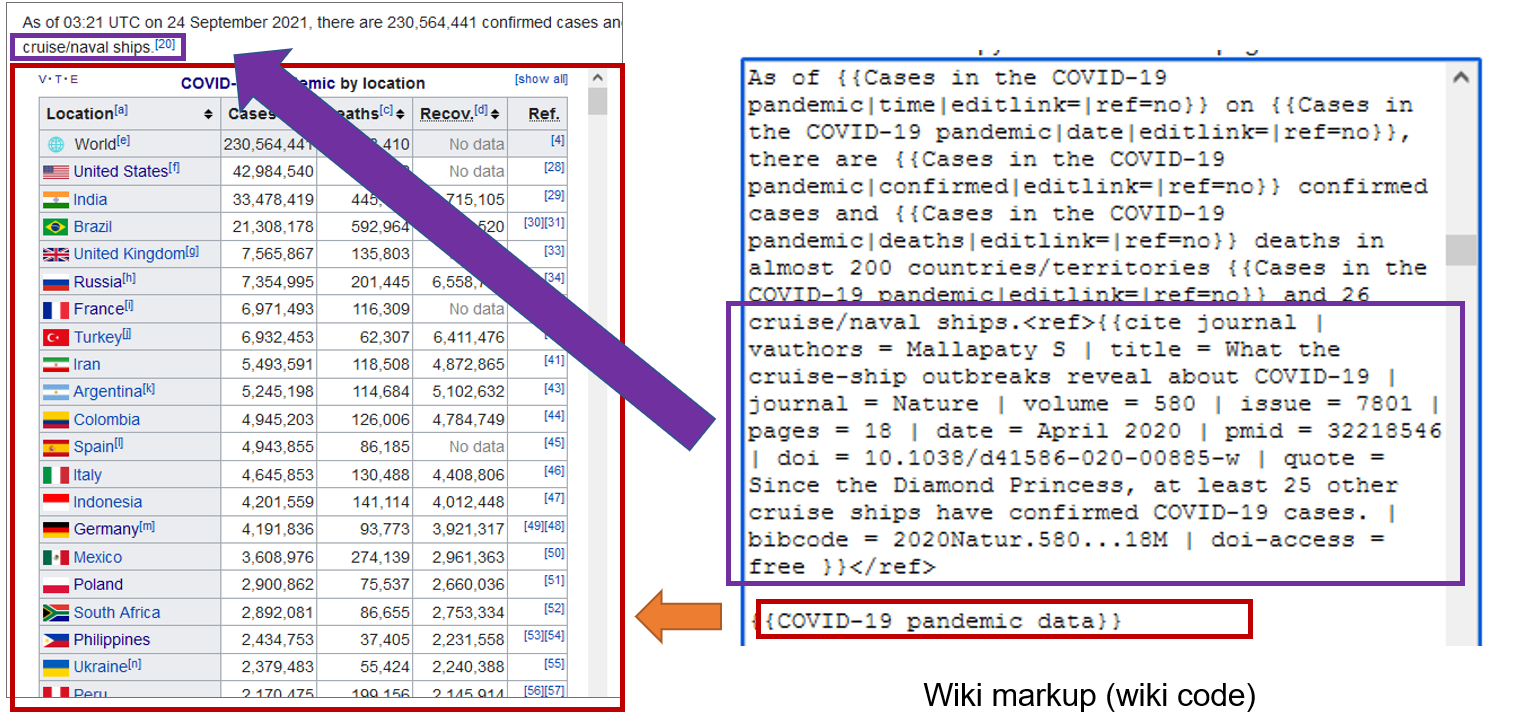}
	\caption{An example of content with references, which was placed using template ''COVID-19 pandemic data'' (as of September 2021). Source: own work based on \cite{url23}}
	\label{fig:wikicode}
\end{figure}

Revision history of Wikipedia articles in dumps or through API is presented in wiki markup. Such markup allows to put special templates, which put on the article content from other pages. Which makes it much harder to get all the references that a Wikipedia reader sees in compiled (final) version of article. An example of such a situation is shown in the figure \ref{fig:wikicode}. As we can see, apart of reference in sentence (which can be relatively easily detected and necessary data can be extracted) there is a template ''COVID-19 pandemic by country and territory'' which generate a table with references in a final (rendered) version of the article ''COVID-19 pandemic by country and territory'' in English Wikipedia \citep{url23}.

If we want to properly analyze content, which Wikipedia readers seen in the past, we need also be able to find appropriate historical content, which is placed by such template. Moreover, to be able to include content from older versions of templates, we need to know all alternative names for such template. For purpose of this study special algorithms were developed, which match the date of article with historical version of such template.

Similarly to previous work \citep{lewoniewski2020} this study used own complex extraction method with some modifications and improvements. In order to detect and extract data from references own algorithm in Python was written. Some of the features of this algorithm are described below. 

References in wiki markup are usually placed between special tags \textit{<ref>…</ref>}. Additionally each reference can be named -  by adding ''name'' parameter to this tag: \textit{<ref name=''...''>...</ref>}. If such reference with name was defined in the selected article, it can be placed elsewhere in the same article using only \textit{<ref name=''...'' />}. So, we can use the same reference several times without providing detailed information about it again. However, there are also other possibilities insert references, that were defined in some place of Wikipedia article. More over, defining of reference with its metadata can be done also in different ways. To do so, Wikipedia authors can also use special citation templates with specific names and parameters set. Some of the templates do not require to put references under \textit{<ref>...</ref>} tag.

If references don't use special template, they usually have URL of source and some optional description (such as title of the external page). If reference use one of special templates, it can have more possibility to describe the source. In such templates on separate parameters you can add information about title, URL, author(s), format, access date, journal, publisher, and others. The set of possible parameters with predefined names depends on language version of Wikipedia. More over, the set of possible parameters depends also on type of source: web page, book, journal, news portal, conference and others. For example, among the most commonly used templates in English Wikipedia are: 'Cite web', 'Cite news', 'Cite book', 'Cite journal', 'NHLE', 'Cite magazine' and others \citep{lewoniewski2020}.

The most important data for this study is URL addresses (external links) of the sources in references of Wikipedia articles. However, sometimes important sources of information are placed not as reference. For example, official site and Twitter account of Polish Ministry of Health are provided as a sources of data for Poland medical cases chart in Wikipedia article about COVID-19 pandemic in Poland \citep{url25}. Such note is placed in the form of sentence below the chart in corresponding template \citep{url26}. Therefore additionally such sources were also extracted for purpose of this research.

\begin{figure}[h]
	\centering
	\includegraphics[width=\linewidth]{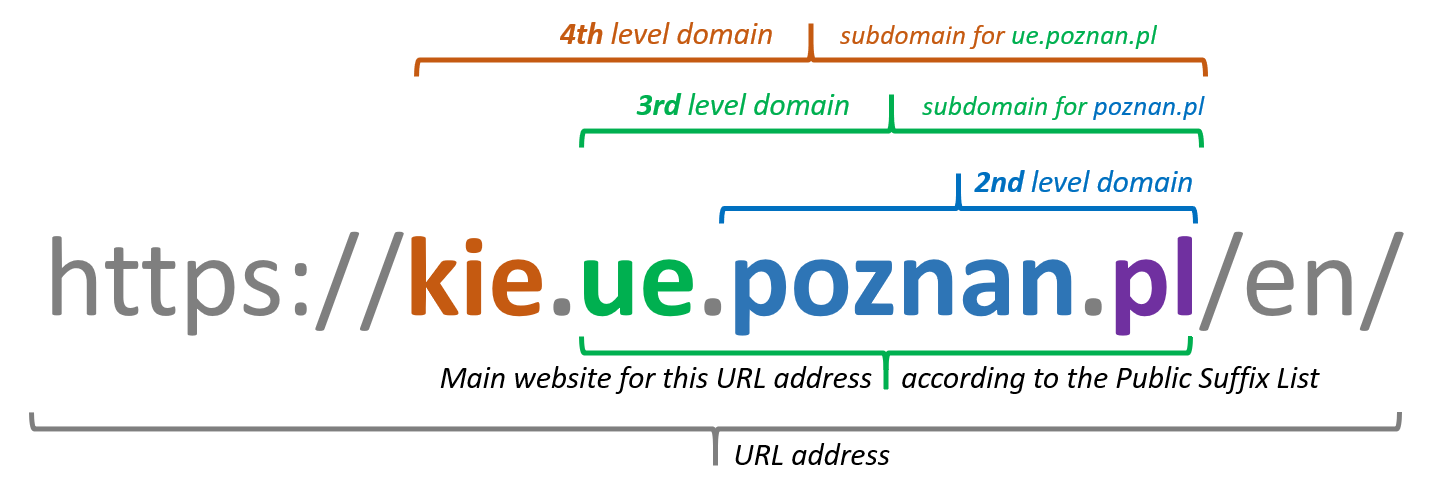}
	\caption{Example of URL address at fourth level domain with indication of main organizational website using PSL}
	\label{fig:urlscheme}
\end{figure}

After extracting external links (URL addressees), we can indicate web domain. However, depending on the web site, it can use different structure of URL addresses. For example, sources can use subdomains for separate topics of news or some organizational unit may post its news on subdomain of main organization. In order to detect which level of domain indicates the source in this study used the Public Suffix List (PSL) - a cross-vendor initiative to provide an accurate list of domain name suffixes \citep{url24}. Example of URL address at fourth level domain with indication of main organizational website using PSL is shown on figure \ref{fig:urlscheme}.

\subsection{Page views data}

Next step is to extract data of page views of each Wikipedia article in each considered language version. To do so, we can use page views dumps or special online tool \citep{url21}. The most detailed dumps files are generated on the fly each our. However, visits from all languages are placed in common files, and even if we need only data about visits of few articles from selected languages, we need to analyze data about all registered visits. This makes it necessary to analyze the data of a relatively large volume. Another way to get page view data is to use dedicated API service \cite{url22}. However each title of the article (and separately each redirect) needs separate request

It is important to note, that in the process of obtaining data on the popularity of specific Wikipedia articles we need to have all alternative names of them. This is because an article may change its name over time and page views data were written under this old name at certain time intervals. 

Usually, when renaming a Wikipedia article, an automatic redirection is created in place of the previous title. Therefore, even if we saved an old URL address of a page in the past, we would easily find a newer version of the same article. In other words, if we analyze all redirects to the article we will able to know previous and alternative names for some subject which in turn will help us to conduct more complete popularity analysis for selected period of time. Information of such redirects can be extracted from Wikimedia dumps or through querying Wikipedia API in each language version.

Table \ref{tab:stats} presents the number of Wikipedia authors and page views for articles related to COVID-19 pandemic in each selected language versions.

\begin{table}[htbp]
	\centering
	\begin{scriptsize}
		\begin{tabular}{|l|r|r|r|r|r|}
			\hline
			\multicolumn{1}{|c|}{\multirow{2}{*}{\textbf{Language}}} & \multicolumn{1}{c|}{\multirow{2}{*}{\textbf{Articles}}} & \multicolumn{2}{c|}{\textbf{Number of authors}}                               & \multicolumn{2}{c|}{\textbf{Number of views}} \\ \cline{3-6} 
			\multicolumn{1}{|c|}{}                                   & \multicolumn{1}{c|}{}                                   & \multicolumn{1}{c|}{\textbf{All}} & \multicolumn{1}{c|}{\textbf{Registered}} & \multicolumn{1}{c|}{\textbf{All}}                    & \multicolumn{1}{c|}{\textbf{Humans}}                    \\ \hline
			ar - Arabic                                              & 371                                                     & 703                               & 456                                       & 6,412,847                                            & 5,344,165                                               \\ \hline
			de - German                                              & 237                                                     & 6,709                             & 2,318                                     & 185,463,741                                          & 34,857,401                                              \\ \hline
			en - English                                             & 547                                                     & 33,462                            & 13,366                                    & 485,458,263                                          & 396,324,156                                             \\ \hline
			es - Spanish                                             & 276                                                     & 4,112                             & 1,460                                     & 31,864,279                                           & 28,911,313                                              \\ \hline
			fr - French                                              & 208                                                     & 4,539                             & 2,017                                     & 33,792,056                                           & 29,691,397                                              \\ \hline
			it - Italian                                             & 184                                                     & 2,512                             & 748                                       & 16,429,153                                           & 10,611,966                                              \\ \hline
			ja - Japanese                                            & 51                                                      & 1,554                             & 663                                       & 8,378,678                                            & 7,440,894                                               \\ \hline
			nl - Dutch                                               & 46                                                      & 773                               & 415                                       & 3,356,003                                            & 3,044,814                                               \\ \hline
			pl - Polish                                              & 41                                                      & 880                               & 454                                       & 5,010,676                                            & 4,036,987                                               \\ \hline
			pt - Portuguese                                          & 255                                                     & 863                               & 455                                       & 8,837,436                                            & 7,430,726                                               \\ \hline
			ru - Russian                                             & 129                                                     & 2,326                             & 1,026                                     & 22,760,150                                           & 21,618,366                                              \\ \hline
			sv - Swedish                                             & 7                                                       & 268                               & 180                                       & 913,751                                              & 719,997                                                 \\ \hline
			uk - Ukrainian                                           & 254                                                     & 410                               & 275                                       & 1,706,842                                            & 909,935                                                 \\ \hline
			vi - Vietnamese                                          & 211                                                     & 1,368                             & 408                                       & 6,439,826                                            & 5,852,113                                               \\ \hline
			zh - Chinese                                             & 218                                                     & 3,542                             & 1,503                                     & 32,725,696                                           & 28,190,101                                              \\ \hline
		\end{tabular}
	\end{scriptsize}
	\caption{Number of Wikipedia authors and page views in the period January 2020 - August 2021 for articles related to COVID-19 pandemic in each selected language versions.}
	\label{tab:stats}
\end{table}

\section{Identifications of important sources on COVID-19 pandemic in Wikipedia}
After extraction of sources and additional metadata (such as page views statistics) for considered articles daily readability scores were counted using three models (described earlier): F, PR, PR2. Next those values were grouped into the mounts based on average. 

With information available separately for each language version of Wikipedia, let's look at the statistics for all and some of them. Please note, that due to limitation of the size in this paper only some of the results were showed and described. Therefore, more detailed statistics are published in supplementary materials to this research on \cite{suppl}.

\begin{figure}[h]
	\centering
	\includegraphics[width=\linewidth]{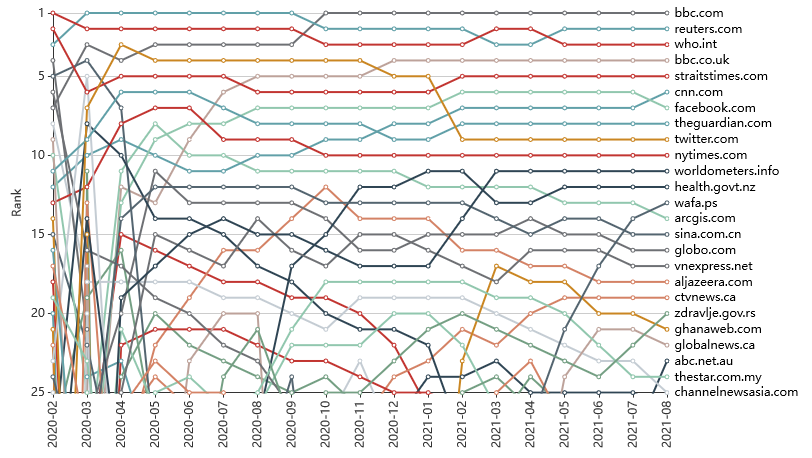}
	\caption{Rank timeline of the most important websites as a source of information on COVID-19 pandemic in 15 languages using F-model. More extended version on \cite{suppl}.}
	\label{fig:wykres-f}
\end{figure}

First,  let's analyze all articles related to COVID-19 pandemic in 15 considered languages (number of articles with their statistics showed in tables \ref{tab:articlescount} and \ref{tab:stats}). The figure \ref{fig:wykres-f} shows Rank trend for the most important sources of information on COVID-19 in 15 language versions using F-model (frequency of the websites). We can see, that among the most frequent sources of COVID-19 pandemic information were: BBC, Reuters, World Health Organization (WHO), The Straits Times, CNN, Facebook, The Guardian, Twitter, The New York Times.

\begin{figure}[h]
	\centering
	\includegraphics[width=\linewidth]{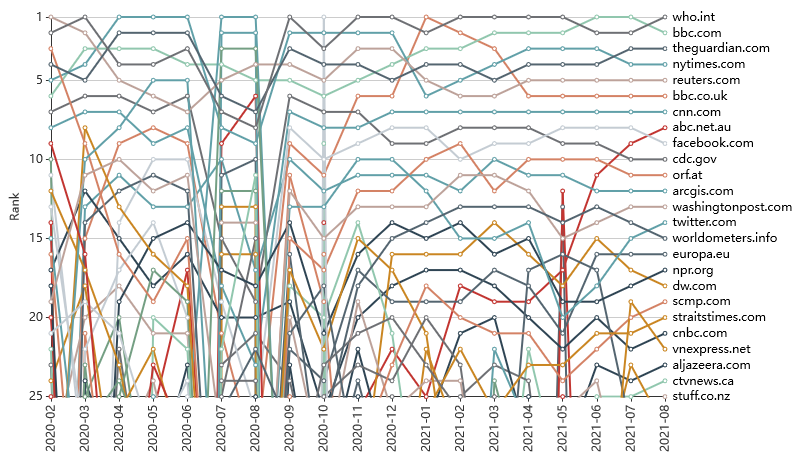}
	\caption{Rank timeline of the most important websites as a source of information on COVID-19 pandemic in 15 language versions using PR-model. Extended version on \cite{suppl}.}
	\label{fig:wykres-pr}
\end{figure}

After considering additionally the popularity of separate Wikipedia articles and the number of references in them, we can observe some changes in the rank timeline. The figure \ref{fig:wykres-pr} shows Rank trend for the most important sources of information on COVID-19 in 15 language versions using PR-model. Comparing to F-model results, here we have some common leaders in website ranking: WHO, BBC, The Guardian, The New York Times, Reuters, CNN, Facebook. Also there are new (comparing to F-model) sources at the top: The Washington Post, Centers for Disease Control and Prevention, Austrian Broadcasting Corporation. Additionally we see, that in case of PR-model we can observe greater volatility of sources ranks during between months.

\begin{figure}[h]
	\centering
	\includegraphics[width=\linewidth]{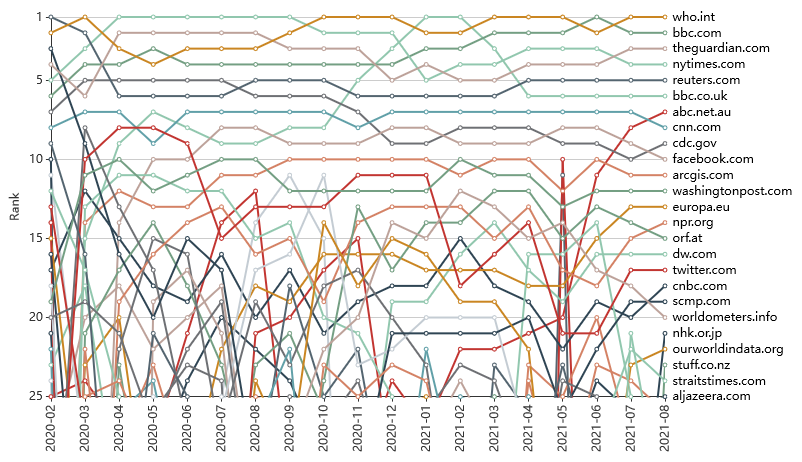}
	\caption{Rank timeline of the most important websites as a source of information on COVID-19 pandemic in 15 languages using PR2-model. More extended version on \cite{suppl}.}
	\label{fig:wykres-pr2}
\end{figure}

If we only count page visits from people (PR2-model), we will find a reduction of volatility, with great similarity to PR-model than to F-model. The figure \ref{fig:wykres-pr2} shows rank trend for the most important sources of information on COVID-19 in 15 languages using PR2-model.

\begin{figure}
	\centering
	\includegraphics[width=\linewidth]{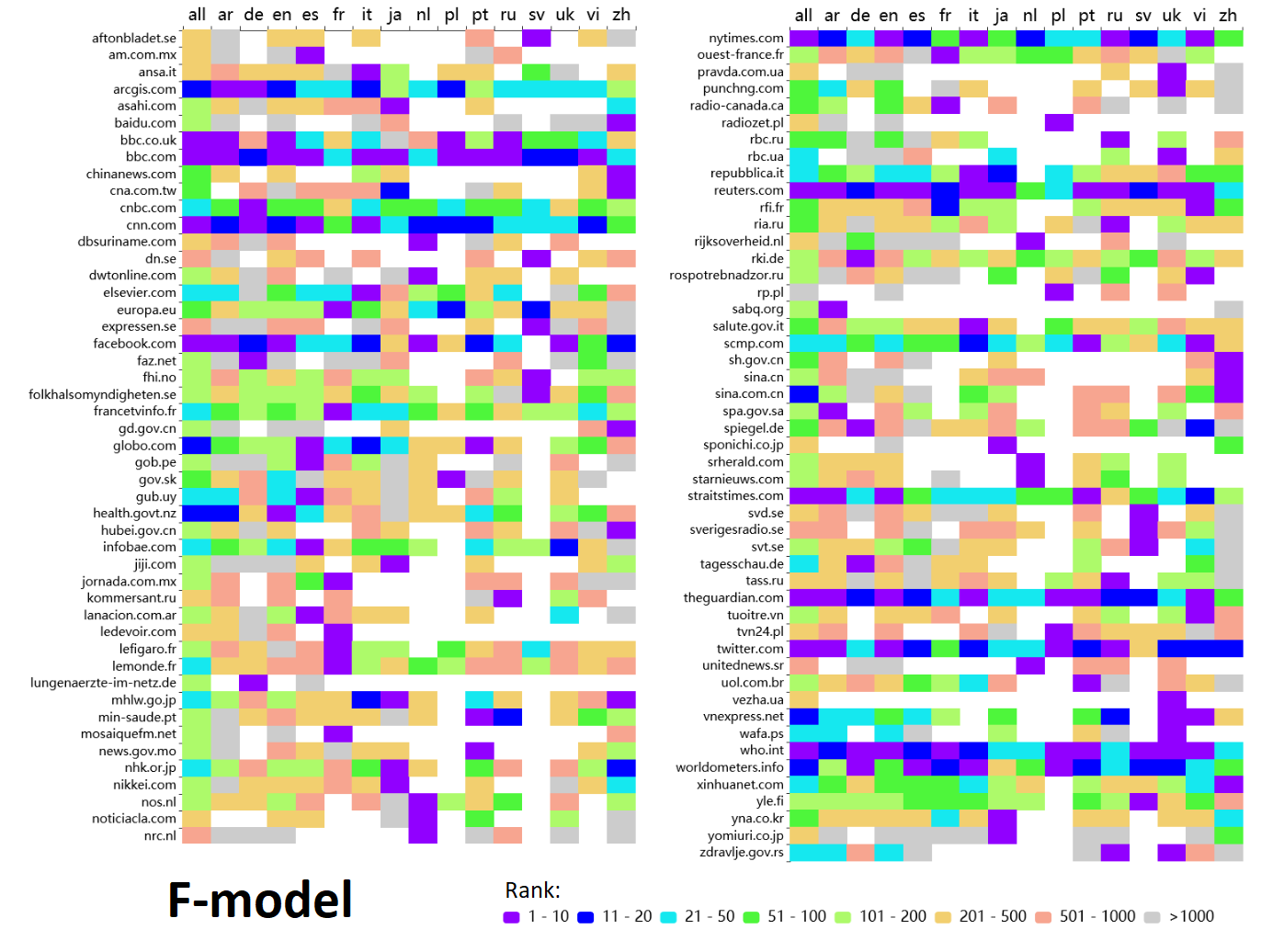}
	\caption{Average ranks of the most important websites as a source of information on COVID-19 pandemic in each of 15 language versions using F-model. Interactive version on \cite{suppl}.}
	\label{fig:heatmapf}
\end{figure}

Let's now compare importance of the some sources between language versions of Wikipedia. To limit the chart size, only sources that appears in top 10 websites at least in one of 15 considered language versions of Wikipedia were selected. Additionally, reliability scores were averaged from each months. Figure \ref{fig:heatmapf} present such comparison as a heat map using ranks of important sources through F-model of each Wikipedia languages.

You can see that among the most important sources of Wikipedia articles on COVID-19 pandemic in high positions at the same time in several language versions can be found such websites as: ArcGIS Online, CNBC, BBC, CNN, The New York Times, Reuters, Twitter, WHO, Worldometer.

\begin{figure}[h]
	\centering
	\includegraphics[width=\linewidth]{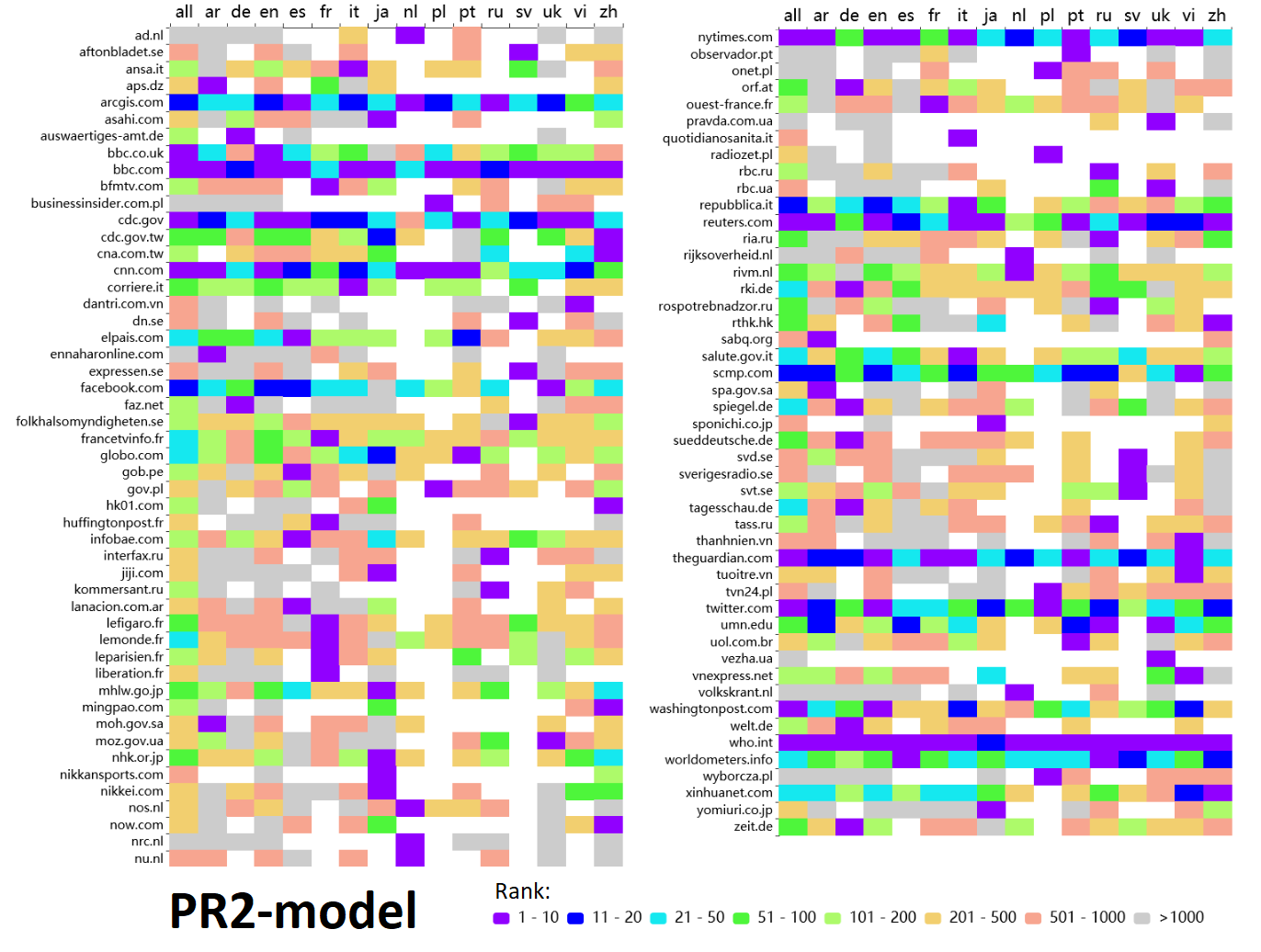}
	\caption{Average ranks of the most important websites as a source of information on COVID-19 pandemic in each of 15 language versions using PR2-model. More on \cite{suppl}.}
	\label{fig:heatmappr2}
\end{figure}

After using PR2-model we can expect some changes in the results. Figure \ref{fig:heatmappr2} shows average ranks of the most important websites as a source of information on COVID-19 pandemic in each of 15 language versions using PR2-model.

Comparing to results from F-model, we can find, that there are some new important sources on the list. For example one of the most important websites on COVID-19 pandemic with high ranks at the same time in several language are: Centers for Disease Control and Prevention, The Guardian, South China Morning Post.

Next subsections described results for some of the languages chapters of Wikipedia. To find names and descriptions of the websites this research used data from semantic databases DBpedia, Wikidata and corresponding Wikipedia articles.

\subsection{Arabic Wikipedia}
\begin{figure}[h]
	\centering
	\includegraphics[width=\linewidth]{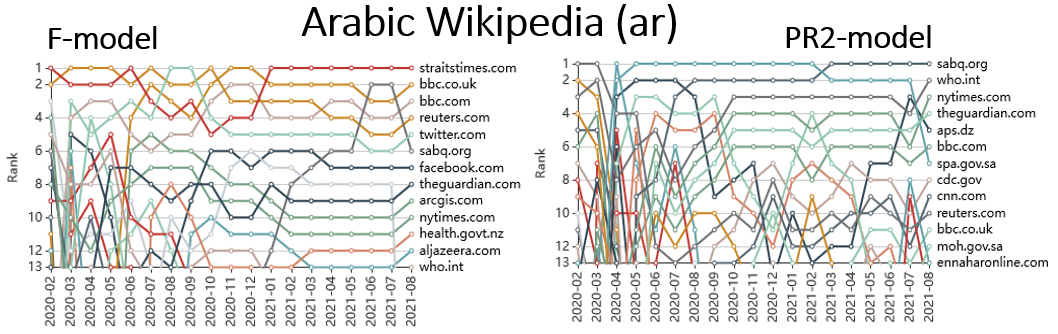}
	\caption{Rank trend for the most important sources of information on COVID-19 in Arabic Wikipedia using F-model and PR2-model. More extended version on \cite{suppl}.}
	\label{fig:wikiar}
\end{figure}

Arabic Wikipedia contained 371 articles on COVID-19 pandemic. In the considered period of time those articles were edited by 703 unique users and were viewed 6.4 million times. The figure \ref{fig:wikiar} presents results of reliability assessment of the websites as sources on COVID-19 pandemic in Arabic chapter of the encyclopedia in each months using F-model and PR2-model. 

One of the most important sources on COVID-19 pandemic in Arabic Wikipedia according to both models in various months were: The New York Times, WHO, BBC, Shahdnow (news website), Reuters, The Guardian.

\subsection{Chinese Wikipedia}
\begin{figure}[h]
	\centering
	\includegraphics[width=\linewidth]{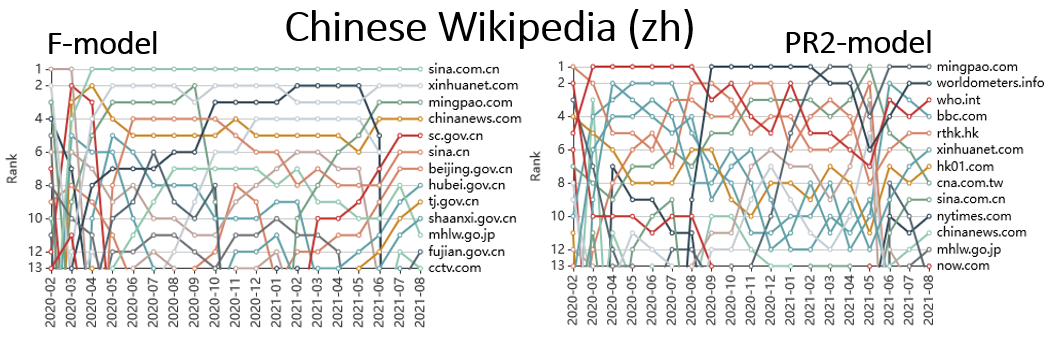}
	\caption{Rank trend for the most important sources of information on COVID-19 in Chinese Wikipedia using F-model and PR2-model. More extended version on \cite{suppl}.}
	\label{fig:wikizh}
\end{figure}

All told, 218 articles on COVID-19 pandemic were found on Chinese Wikipedia. In the considered period of time those articles were edited by 3,542 unique users and were viewed 32.7 million times. The figure \ref{fig:wikizh} presents results of reliability assessment of the websites as sources on COVID-19 pandemic in Chinese chapter of the encyclopedia in each months using F-model and PR2-model. 

One of the most important sources on COVID-19 pandemic in Chinese Wikipedia according to both models in various months were: Ming Pao (newspaper), WHO, China News Service (news agency), Sina (infotainment portal), Worldometer. 

\subsection{Dutch Wikipedia}
\begin{figure}[h]
	\centering
	\includegraphics[width=\linewidth]{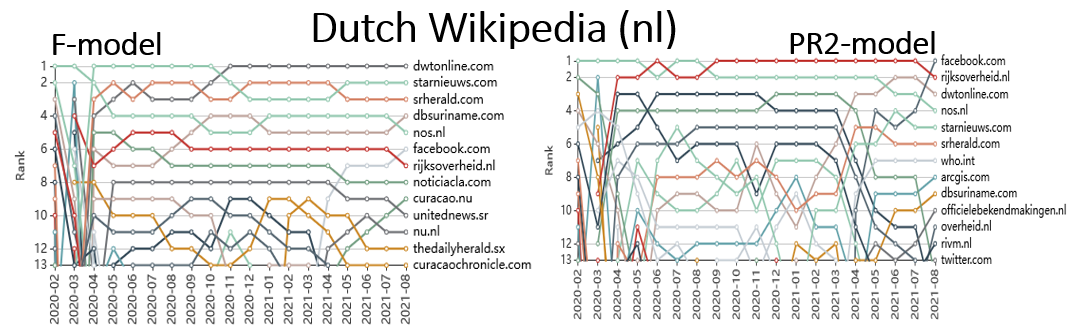}
	\caption{Rank trend for the most important sources of information on COVID-19 in Dutch Wikipedia using F-model and PR2-model. More extended version on \cite{suppl}.}
	\label{fig:wikinl}
\end{figure}

There were 46 articles on COVID-19 pandemic in Dutch Wikipedia. In the considered period of time those articles were edited by 773 unique users and were viewed 3.4 million times. The figure \ref{fig:wikinl} presents results of reliability assessment of the websites as sources on COVID-19 pandemic in Dutch chapter of the encyclopedia in each months using F-model and PR2-model. 

One of the most important sources on COVID-19 pandemic in Dutch Wikipedia according to both models in various months were: De Ware Tijd (daily newspapers), Starnieuws (news website), Suriname Herald (news website), Nederlandse Omroep Stichting (service broadcaster and news network), Netherlands central government, WHO. 

\subsection{English Wikipedia}
\begin{figure}[h]
	\centering
	\includegraphics[width=\linewidth]{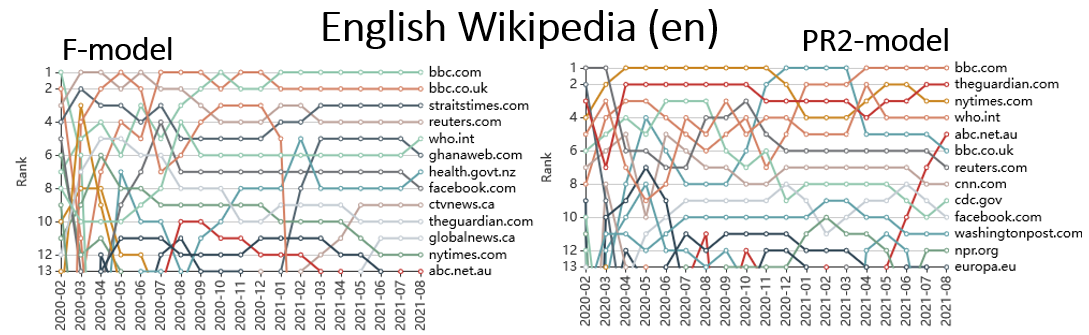}
	\caption{Rank trend for the most important sources of information on COVID-19 in English Wikipedia using F-model and PR2-model. More extended version on \cite{suppl}.}
	\label{fig:wikien}
\end{figure}

Over all, 547 articles related to COVID-19 pandemic were found on English Wikipedia. In the considered period of time those articles were edited by 33,462 unique users and were viewed 485.5 million times. The figure \ref{fig:wikien} presents results of reliability assessment of the websites as sources on COVID-19 pandemic in English chapter of the encyclopedia in each months using F-model and PR2-model. 

One of the most important sources on COVID-19 pandemic in English Wikipedia according to both models in various months were: BBC, WHO, The Guardian, Reuters, The New York Times. 

\subsection{French Wikipedia}
\begin{figure}[h]
	\centering
	\includegraphics[width=\linewidth]{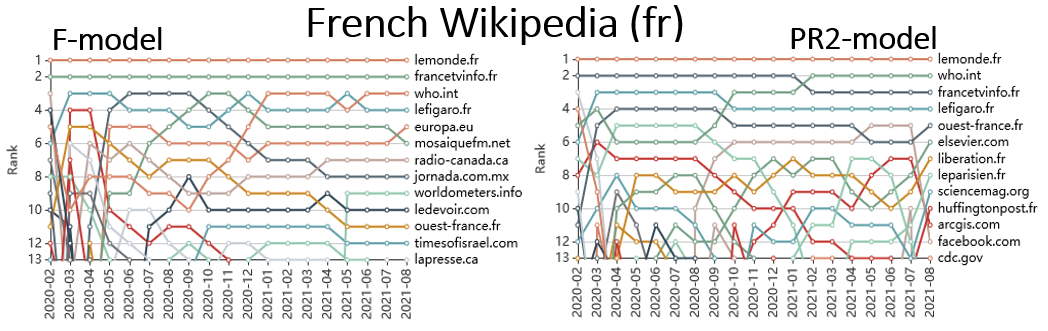}
	\caption{Rank trend for the most important sources of information on COVID-19 in French Wikipedia using F-model and PR2-model. More extended version on \cite{suppl}.}
	\label{fig:wikifr}
\end{figure}

French Wikipedia contained 208 articles on COVID-19 pandemic. In the considered period of time those articles were edited by 4,539 unique users and were viewed 33.8 million times. The figure \ref{fig:wikifr} presents results of reliability assessment of the websites as sources on COVID-19 pandemic in French chapter of the encyclopedia in each months using F-model and PR2-model. 

One of the most important sources on COVID-19 pandemic in French Wikipedia according to both models in various months were: Le Monde (daily afternoon newspaper), France Info (news channel), WHO, Le Figaro (daily morning newspaper), Ouest-France (newspaper). 

\subsection{German Wikipedia}
\begin{figure}[h]
	\centering
	\includegraphics[width=\linewidth]{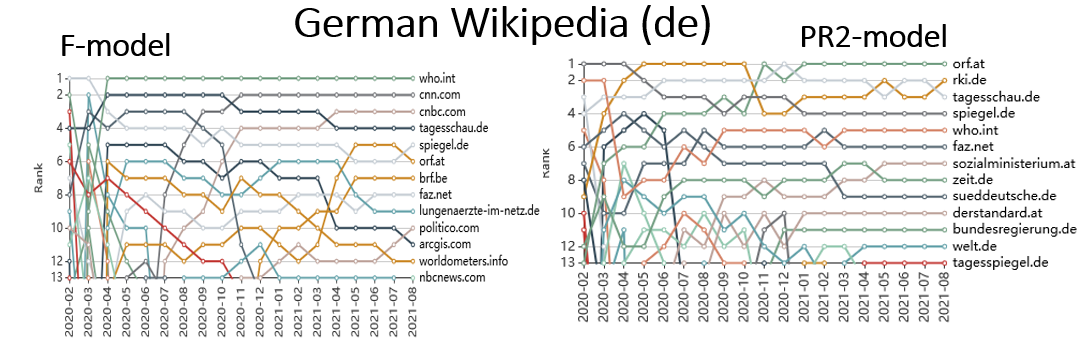}
	\caption{Rank trend for the most important sources of information on COVID-19 in German Wikipedia using F-model and PR2-model. More extended version on \cite{suppl}.}
	\label{fig:wikide}
\end{figure}

237 articles on COVID-19 pandemic were found on German Wikipedia. In the considered period of time those articles were edited by 6,709 unique users and were viewed 185.5 million times. The figure \ref{fig:wikide} presents results of reliability assessment of the websites as sources on COVID-19 pandemic in German chapter of the encyclopedia in each months using F-model and PR2-model. 

One of the most important sources on COVID-19 pandemic in German Wikipedia according to both models in various months were: Tagesschau (television news service), Der Spiegel (news magazine and news website), WHO, Frankfurter Allgemeine Zeitung (newspaper). 

\subsection{Italian Wikipedia}
\begin{figure}[h]
	\centering
	\includegraphics[width=\linewidth]{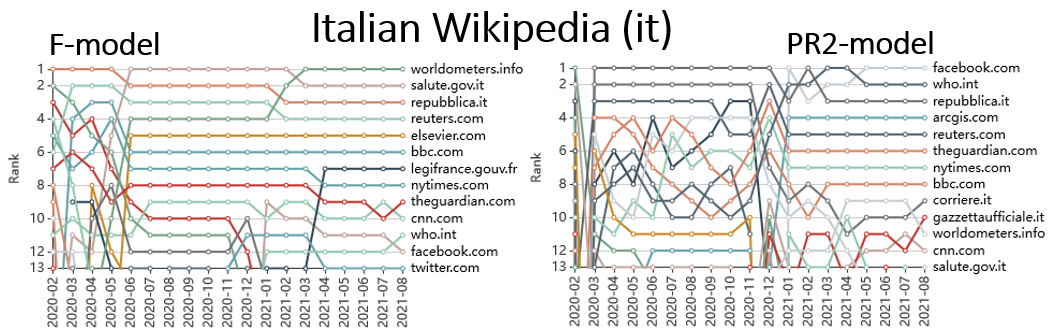}
	\caption{Rank trend for the most important sources of information on COVID-19 in Italian Wikipedia using F-model and PR2-model. More extended version on \cite{suppl}.}
	\label{fig:wikiit}
\end{figure}

There were 184 articles on COVID-19 pandemic in Italian Wikipedia. In the considered period of time those articles were edited by 2,512 unique users and were viewed 16.4 million times. The figure \ref{fig:wikiit} presents results of reliability assessment of the websites as a source on COVID-19 pandemic in Italian chapter of the encyclopedia in each months using F-model and PR2-model.

\subsection{Japanese Wikipedia}
\begin{figure}[h]
	\centering
	\includegraphics[width=\linewidth]{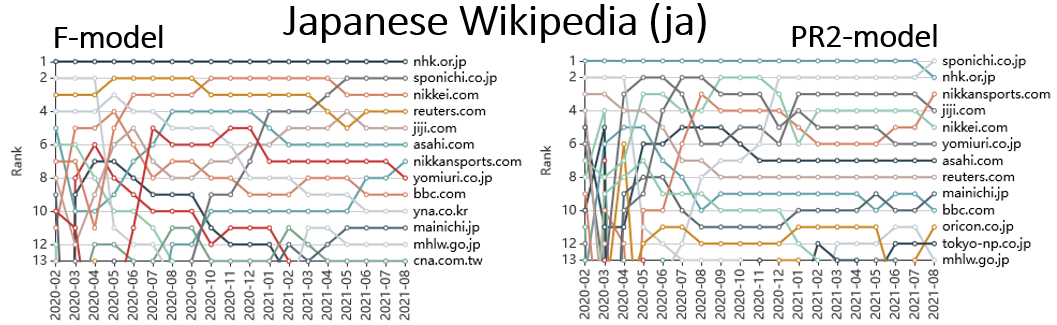}
	\caption{Rank trend for the most important sources of information on COVID-19 in Japanese Wikipedia using F-model and PR2-model. More extended version on \cite{suppl}.}
	\label{fig:wikija}
\end{figure}

Japanese Wikipedia contained 51 articles on COVID-19 pandemic. In the considered period of time those articles were edited by 1,554 unique users and were viewed 8.4 million times. The figure \ref{fig:wikija} presents results of reliability assessment of the websites as a source on COVID-19 pandemic in Japanese chapter of the encyclopedia in each months using F-model and PR2-model.

One of the most important sources on COVID-19 pandemic in Japanese Wikipedia according to both models in various months were: NHK (public broadcaster), Sports Nippon (daily sports newspaper), The Nikkei (daily newspaper), Jiji Press (news agency), The Asahi Shimbun (daily newspaper), Nikkan Sports (daily newspaper), Reuters, BBC.

\subsection{Polish Wikipedia}
\begin{figure}[h]
	\centering
	\includegraphics[width=\linewidth]{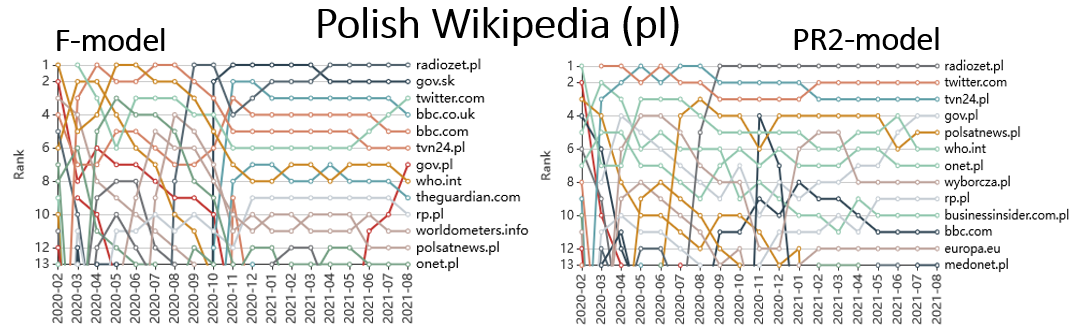}
	\caption{Rank trend for the most important sources of information on COVID-19 in Polish Wikipedia using F-model and PR2-model. More extended version on \cite{suppl}.}
	\label{fig:wikipl}
\end{figure}

All told, 41 articles on COVID-19 pandemic were found on Polish Wikipedia. In the considered period of time those articles were edited by 880 unique users and were viewed 5 million times. The figure \ref{fig:wikipl} presents results of reliability assessment of the websites as a source on COVID-19 pandemic in Polish chapter of the encyclopedia in each months using F-model and PR2-model.

One of the most important sources on COVID-19 pandemic in Polish Wikipedia according to both models in various months were: Radio ZET (radio station), 
Twitter, TVN24 (news channel), GOV.PL (official service of Poland), Onet (web portal), Rzeczpospolita (daily newspaper), Polsat News (news channel), WHO, Gazeta Wyborcza (daily newspaper).

\subsection{Portuguese Wikipedia}
\begin{figure}[h]
	\centering
	\includegraphics[width=\linewidth]{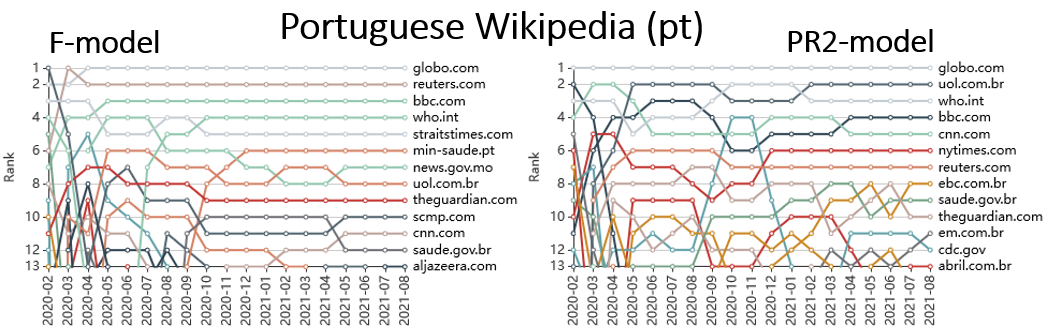}
	\caption{Rank trend for the most important sources of information on COVID-19 in Portuguese Wikipedia using F-model and PR2-model. More extended version on \cite{suppl}.}
	\label{fig:wikipt}
\end{figure}

There were 255 articles on COVID-19 pandemic in Portuguese Wikipedia. In the considered period of time those articles were edited by 863 unique users and were viewed 8.8 million times. The figure \ref{fig:wikipt} presents results of reliability assessment of the websites as a source on COVID-19 pandemic in Portuguese chapter of the encyclopedia in each months using F-model and PR2-model.

One of the most important sources on COVID-19 pandemic in Portuguese Wikipedia according to both models in various months were: Globo (web portal), WHO, BBC, Reuters, Universo Online (web portal), Governo do Brasil (official web portal).

\subsection{Russian Wikipedia}
\begin{figure}[h]
	\centering
	\includegraphics[width=\linewidth]{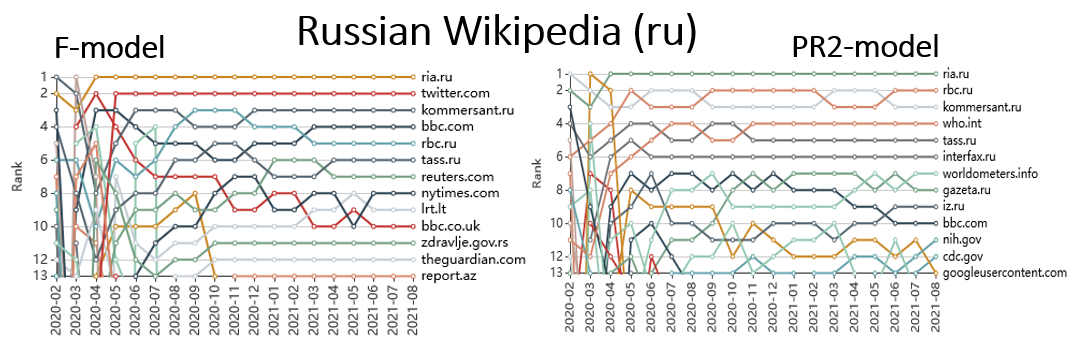}
	\caption{Rank trend for the most important sources of information on COVID-19 in Russian Wikipedia using F-model and PR2-model. More extended version on \cite{suppl}.}
	\label{fig:wikiru}
\end{figure}

Russian Wikipedia contained 129 articles on COVID-19 pandemic. In the considered period of time those articles were edited by 2,326 unique users and were viewed 22.8 million times. The figure \ref{fig:wikiru} presents results of reliability assessment of the websites as a source on COVID-19 pandemic in Russian chapter of the encyclopedia in each months using F-model and PR2-model.

One of the most important sources on COVID-19 pandemic in Russian Wikipedia according to both models in various months were: RIA Novosti (news agency), RBK (news web-portal and business newspaper), Kommersant (daily newspaper), TASS (news agency), Interfax (news agency).

\subsection{Spanish Wikipedia}
\begin{figure}[h]
	\centering
	\includegraphics[width=\linewidth]{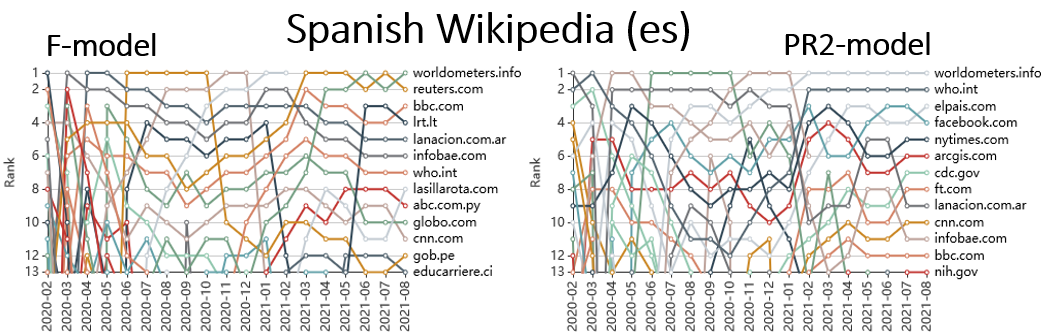}
	\caption{Rank trend for the most important sources of information on COVID-19 in Spanish Wikipedia using F-model and PR2-model. More extended version on \cite{suppl}.}
	\label{fig:wikies}
\end{figure}

276 articles on COVID-19 pandemic were found on Spanish Wikipedia. In the considered period of time those articles were edited by 4,112 unique users and were viewed 31.9 million times. The figure \ref{fig:wikies} presents results of reliability assessment of the websites as a source on COVID-19 pandemic in Spanish chapter of the encyclopedia in each months using F-model and PR2-model.

One of the most important sources on COVID-19 pandemic in Spanish Wikipedia according to both models in various months were: Worldometer, El País (daily newspaper), BBC, Infobae (news website).

\subsection{Swedish Wikipedia}
\begin{figure}[h]
	\centering
	\includegraphics[width=\linewidth]{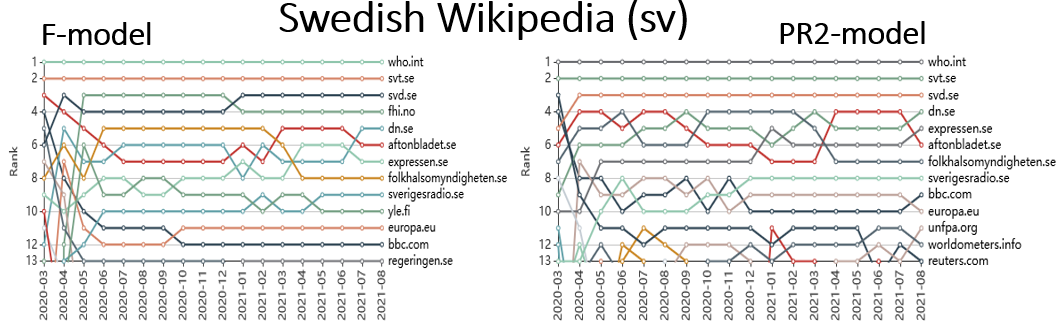}
	\caption{Rank trend for the most important sources of information on COVID-19 in Swedish Wikipedia using F-model and PR2-model. More extended version on \cite{suppl}.}
	\label{fig:wikisv}
\end{figure}

There were 7 articles on COVID-19 pandemic in Swedish Wikipedia. In the considered period of time those articles were edited by 180 unique users and were viewed 913 thousand times. The figure \ref{fig:wikisv} presents results of reliability assessment of the websites as sources on COVID-19 pandemic in Swedish chapter of the encyclopedia in each months using F-model and PR2-model. 

One of the most important sources on COVID-19 pandemic in Swedish Wikipedia according to both models in various months were: WHO, Sveriges Television (television broadcaster), Svenska Dagbladet (daily newspaper), Dagens Nyheter (daily newspaper), Aftonbladet (daily newspapers), Public Health Agency of Sweden, Sveriges Radio (radio broadcaster). 

\subsection{Ukrainian Wikipedia}
\begin{figure}[h]
	\centering
	\includegraphics[width=\linewidth]{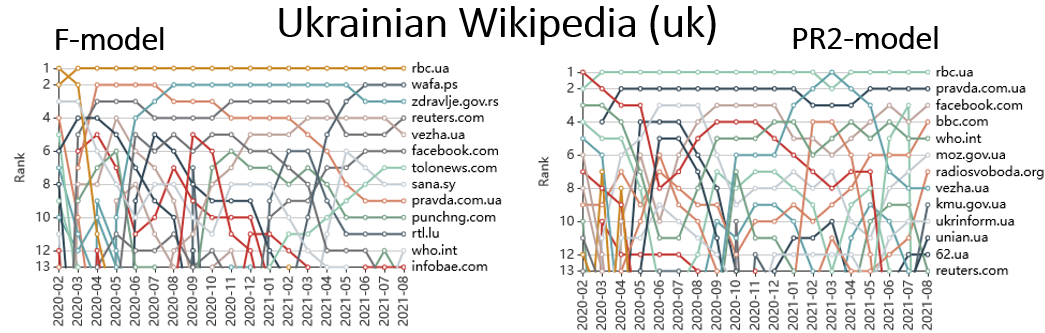}
	\caption{Rank trend for the most important sources of information on COVID-19 in Ukrainian Wikipedia using F-model and PR2-model. More extended version on \cite{suppl}.}
	\label{fig:wikiuk}
\end{figure}

Ukrainian Wikipedia contained 254 articles on COVID-19 pandemic. In the considered period of time those articles were edited by 410 unique users and were viewed 1.7 million times. The figure \ref{fig:wikiuk} presents results of reliability assessment of the websites as a source on COVID-19 pandemic in Ukrainian chapter of the encyclopedia in each months using F-model and PR2-model.

One of the most important sources on COVID-19 pandemic in Ukrainian Wikipedia according to both models in various months were: RBC Ukraine (news agency), Ukrayinska Pravda (online newspaper), Reuters, WHO, Vezha (information web portal)

\subsection{Vietnamese Wikipedia}
\begin{figure}[h]
	\centering
	\includegraphics[width=\linewidth]{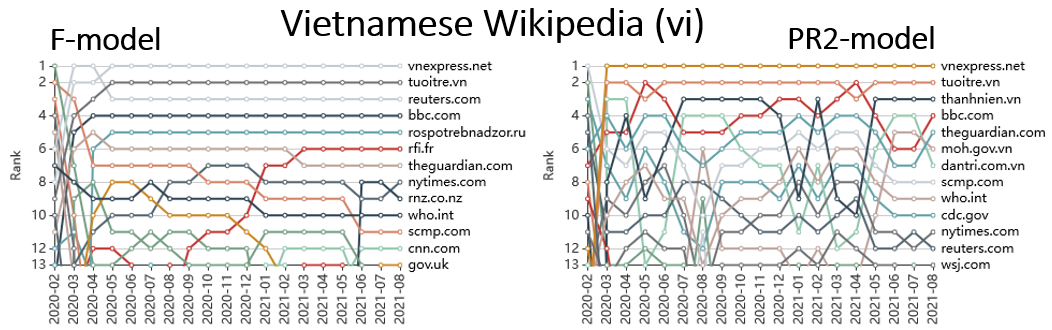}
	\caption{Rank trend for the most important sources of information on COVID-19 in Vietnamese Wikipedia using F-model and PR2-model. More extended version on \cite{suppl}.}
	\label{fig:wikivi}
\end{figure}

Over all, 211 articles on COVID-19 pandemic were found on Vietnamese Wikipedia. In the considered period of time those articles were edited by 1,368 unique users and were viewed 6.4 million times. The figure \ref{fig:wikivi} presents results of reliability assessment of the websites as sources on COVID-19 pandemic in Vietnamese chapter of the encyclopedia in each months using F-model and PR2-model. 

One of the most important sources on COVID-19 pandemic in Vietnamese Wikipedia according to both models in various months were: VnExpress (newspaper), Tuoi Tre (daily newspaper), BBC, The Guardian, WHO, Ministry of Health (Vietnam), South China Morning Post (newspaper). 

\section{Conclusion and future work}

In this work, methods of selecting Wikipedia articles on the COVID-19 pandemic in different languages, as well as ways of extracting source information, along with assessing reliability were presented. In particular, three methods for identifying articles using Wikipedia category, Wikidata and DBpedia were shown and explained.

The main focus of the study was on assessing Wikipedia sources during a specific time period and analyzing the rank timeline in each of 15 language chapters of the online encyclopedia. The results show that reliability models that use data about the popularity of Wikipedia articles (PR-model and PR2-model) are able to find important sources on the COVID-19 topic in separate language versions of Wikipedia.

Reliability assessment of the sources on a selected topic can help to improve models for quality assessment of information in Wikipedia and other websites. Such estimation can be especially useful in assessing conflict statements between language versions of Wikipedia or to enrich it with information of the best quality. Additionally, the presented method can help Wikipedia authors by suggesting reliable sources for selected topics and statements in each language version separately.

Future work will focused on extending reliability models. One of the directions is to develop ways of weighting the importance of a reference based on its position within a Wikipedia article. Another promising direction is to include different measures related to the reputation of Wikipedia authors, protection of the articles, topic similarity and others.

\section{Acknowledgments}

The study was conducted within the research project Economics in the face of the New Economy financed within the Regional Initiative for Excellence programme of the Minister of Science and Higher Education of Poland, years 2019-2022, grant no. 004/RID/2018/19, financing 3,000,000 PLN.

\bibliographystyle{ACM-Reference-Format}
\bibliography{literature}


\end{document}